\documentclass[12pt]{article}
\usepackage{epsf}
\usepackage{graphicx}
\usepackage{amssymb}
\usepackage{bbm}

\def\d{{\mathrm{d}}}

\def\u{u}
\def\v{v}

\def\cons{{\chi}}

\date{\today}

\def\C{{\mathbb{C}}}
\def\T{{\mathbb{T}}}
\def\R{{\mathbb{R}}}
\def\Z{{\mathbb{Z}}}

\def\Id{{\mathbbm{1}}}

\def\D{{\mathrm{D} }}

\def\ha{\frac{1}{2}}
\def\iha{\frac{i}{2}}

\begin{document}

\begin{titlepage}

\title{Quark zero modes in intersecting center vortex
gauge fields\footnote{supported by DFG under grant-No. DFG-Re
856/4-1 and DFG 436 RUS 113/477/4}}

\author{H.~Reinhardt, O.~Schr\"oder, T.~Tok\\
\vspace{0.5cm}
University of T\"ubingen\\
V.~Ch.~Zhukovsky\\
Moscow State University}

\maketitle

\vspace{2cm}

\normalsize

\begin{abstract} 
The zero modes of the Dirac operator in the background of  
center vortex gauge field configurations in $\R^2$ and $\R^4$ 
are examined. If the net flux in $D=2$ is larger than $1$ 
we obtain normalizable zero modes which are mainly localized at 
the vortices. In $D=4$ quasi-normalizable zero modes exist 
for intersecting flat vortex sheets with the Pontryagin 
index equal to $2$. These zero modes are mainly localized at the vortex
intersection points, which carry a topological charge of $\pm \ha$. 
To circumvent the problem of normalizability the space-time manifold 
is chosen to be the (compact) torus $\T^2$ and $\T^4$, respectively. 
According to the index theorem there are normalizable zero modes 
on $\T^2$ if the net flux is non-zero. These  zero modes are localized
at the vortices. On $\T^4$ zero modes exist for a non-vanishing 
Pontryagin index. As in $\R^4$ these zero modes are localized at the 
vortex intersection points.
\end{abstract}



\end{titlepage}

\section{Introduction}

There are two fundamental, principally non-perturbative
phenomena of the strong interaction which should be explained by 
QCD: confinement and chiral symmetry
breaking. Soon after the advent
of QCD chiral symmetry breaking has been explained by assuming
that the QCD vacuum contains an ensemble of instantons and 
anti-instantons \cite{Callan-78}. 
Lattice simulations have, however, shown that 
instantons cannot be responsible for confinement since they
account  for only about ten percent of the string tension
\cite{DeGrand-97,Chen-98,Fukushima-00}.
On the other hand, recent lattice calculations have produced 
mounting evidence that confinement is due to the condensation 
of center vortices in the Yang-Mills vacuum 
\cite{DelDebbio-97,Forcrand-99-1}.
Since on the lattice the
confinement phase transition is observed to occur at the same
temperature \cite{Karsch-99}
at which chiral symmetry is restored, one expects that chiral
symmetry breaking and confinement are triggered by the same
mechanism. 

Since instantons cannot explain confinement one
wonders whether center vortices are also capable of producing
chiral symmetry breaking in addition to confinement. Indeed, it
has been shown that in a Yang-Mills ensemble devoid of center
vortices the relevant order parameter, the
quark condensate vanishes \cite{Forcrand-99-1}.
This does, however, not yet mean
that center vortices produce chiral symmetry breaking in the QCD
vacuum, since the field-configurations which produce it 
could be tied or attached to center vortices
and simultaneously removed with the latter. It is therefore
still an open question whether center vortices produce chiral
symmetry breaking and what the underlying mechanism behind
this is. 

In order to get an idea how chiral symmetry breaking could be
produced by center vortices let us first recall how it
arises in the instanton picture of the QCD
vacuum. In this picture the zero modes of quarks in the
instanton background play a crucial role. These zero modes occur
due to  the topological charge of the instantons and are
localized near their topological charge density 
\cite{Feuerstein-97,Ilgenfritz-98a,Ilgenfritz-98b}.

In an instanton-anti-instanton ensemble the localized zero modes of the
individual instantons spread out in space and form a continuum of quasi
zero modes which by the Banks-Casher relation give rise to a non-zero
quark condensate \cite{Zahed}.
In fact, independent of the instanton picture lattice 
calculations show a strong correlation between topological 
charge density and the quark condensate \cite{Ivanenko-97,Sakuler-99}.
This correlation is mainly due to the zero
modes which exist in topologically non-trivial gauge fields with
non-vanishing Pontryagin index due to the Atiyah-Singer index 
theorem \cite{atiyah} and which are localized at topological charge.

In center vortex field configurations topological charge is
concentrated at the intersection points and other singular points
like twisting points \cite{Engelhardt-00-1}.
Near these singular points we expect localization of quark zero 
modes which could play a similar role in the explanation of 
chiral symmetry breaking in the vortex picture as they do 
in the instanton picture. In a first attempt towards an 
understanding of chiral symmetry breaking in the vortex picture, 
in the present paper we study the quark zero modes in
intersecting center vortex background fields. A more rigorous
understanding of the mechanism of chiral symmetry breaking in
the vortex picture is presently under investigation.

We examine zero modes of the Dirac operator in vortex backgrounds 
in two and four dimensions for different Euclidean space-time topologies. 
In sections \ref{modes-d2} and \ref{modes-d4} fermionic zero
modes in the background of vortices in $\R^2$ and 
intersecting flat vortices in $\R^4$ are studied 
and their relation to solutions found by other authors
\cite{Ma-86,Sitenko-96} is shortly discussed.
In sections \ref{modes-t2} and \ref{modes-t4} this analysis is repeated 
for space-time given by $\T^2$ and $\T^4$, respectively.

\section{Fermionic zero modes in non-intersecting center vortex 
fields}
\label{modes-d2}

In $D=4$ center vortices represent closed two-dimensional flux sheets.
We are interested here in the quark modes in the background of such
vortex sheets. Locally a center vortex sheet represents a
two-dimensional plane. For simplicity we will consider in the following
flat (planar) vortex sheets. In this case we have translational
invariance parallel to the vortex sheets and the solution of the Dirac
equation reduces to the two-dimensional problem
in the plane defined by the 2 directions perpendicular to the vortex
sheet. In this plane the vortex appears as intersection point. 
We will first study the Dirac equation in this plane, i.e.~in $D=2$.

We will consider the zero modes of the Dirac operator in 
the background of Abelian gauge potentials representing Dirac
strings and center vortices. 
We can consider these gauge potentials as living in the
Cartan sub-algebra of a $SU(2)$ gauge group. Having this in mind we
say that a gauge potential describes a center vortex or Dirac string 
at a point $z_0$, if the Wilson loop around $z_0$ is $-\Id$ or $+\Id$,
respectively. Equivalently, the magnetic flux carried by a center 
vortex or a Dirac string 
is given by $ \Phi_{center} = ( m+1/2 ) \, , \, m \in \Z$ 
or $ \Phi_{Dirac} = m \, , \, m \in \Z$, respectively\footnote{Our
conventions are summarized in Appendix \ref{conventions}.}.

The solution of the Dirac equation in two dimensions can 
be related to the theory of functions of a complex variable. 
We introduce the complex variable $z$ by
\begin{equation}
z = x + i y \, , \quad \bar z = x - i y \, , \quad 
\partial_z = 
\frac{1}{2} \left( \partial_x - i \partial_y \right) \, , \quad
\partial_{\bar z} = 
\frac{1}{2} \left( \partial_x + i \partial_y  \right) \, 
\end{equation}
and a complex notation for the gauge potential. We define$^1$
\begin{equation}
\label{2.2}
A_z := \frac{1}{2} \left( A_x - i A_y \right) \, , 
\quad A_{\bar z} := \frac{1}{2} \left( A_x + i A_y \right) 
= - \overline{A_z} \, ,
\quad A_x = 2i \Im A_z \, , \quad A_y = 2 i \Re A_z \, ,
\end{equation}
where $\Im A_z$ is the imaginary part and $\Re A_z$ is the real part of
$A_z$.
The Dirac equation$^1$ 
\begin{equation}
\label{Dirac-equation}
i \gamma_\mu \D_\mu \psi 
= \lambda  \psi \, , \quad 
\psi = 
\left( \begin{array}{c} \psi_1 \\ \psi_2 \end{array} \right)
\, , \quad \D_\mu = \partial_\mu + A_\mu  
\end{equation}
in spinor components reads
\begin{eqnarray}
2 i \left( \partial_{z} + A_z \right) \psi_2 
&=& \lambda \psi_1 \, , \\
2 i \left( \partial_{\bar z} + A_{\bar z} \right) \psi_1 
&=& \lambda \psi_2 \, .
\end{eqnarray}

The most simple case is the free Dirac equation, i.e.~$A_z = 0$. 
Zero modes ($\lambda =0$) of the free Dirac equation are obviously 
given by analytic functions $\psi_1$ and anti-analytic functions 
$\psi_2$. A normalizable zero mode has to go to zero at infinity. 
But every (anti-)analytic function without singularities which goes 
to zero at infinity has to be zero everywhere. This means there are no
{\it smooth} zero modes. In principle the zero mode may have 
singularities, in 
this case poles. But a pole in the zero mode is a non-integrable
singularity and the zero mode would be not normalizable.
Hence, there are no normalizable zero modes for the free Dirac equation.

Next we consider the Dirac field in the background of 
a straight plane center vortex. Its gauge potential reads
\begin{equation}
A_x = - y \frac{f(r^2)}{r^2} \frac{i}{2} \, , \quad 
A_y = x  \frac{f(r^2)}{r^2} \frac{i}{2} \, , \quad 
r^2 = x^2 + y^2 = z \bar z \, .
\end{equation} 
Here we introduced the profile function 
$f(r^2)$ which fulfills $f(0) = 0$ and $f(r^2) \to 1$ as $r
\to \infty$.
The complexified gauge potential is then given by
\begin{equation}
\label{complexified}
A_z := \frac{1}{2} \left( A_x - i A_y \right)
= \frac{f(r^2)}{4 r^2} \left( - i y + x \right)
= \frac{f(r^2)}{4 r^2} \left( \bar z \right) \, .
\end{equation}
Introducing the function
\begin{equation}
\phi(x) := \int_1^x \frac{f(x')}{2 x'} \d x'
\end{equation}
the gauge potential $A_z$ can simply be written as
\begin{equation}
A_z = \frac{1}{2} \partial_z \phi( z \bar z) \, .
\end{equation}
Inserting the gauge potential into the Dirac equation 
we obtain the differential equations
\begin{eqnarray}
\label{dirac1}
2 i \left( \partial_{z} + 
\ha \partial_z \phi( z \bar z)  \right) 
\psi_2 &=& \lambda \psi_1 \, , \\
\label{dirac2}
2 i \left( \partial_{\bar z} -
\ha \partial_{\bar z} \phi( z \bar z) \right) 
\psi_1 &=& \lambda \psi_2 \, .
\end{eqnarray}

We are mainly interested in the zero modes $\lambda = 0$. Furthermore,
let us first consider an idealized vortex with $f(r^2) \equiv 1$ (this
function obviously does not have the properties of a profile function,
because $f(0) = 1 \neq 0$).
For $f(r^2) = 1$ the function $\phi$ becomes 
$\phi(r^2) = \ha \log ( r^2 )$, which is - up to a factor - 
the Greens function of the Laplace operator in two dimensions. 
For $\lambda = 0$ and $\phi(r^2) = \ha \log (r^2)$ the 
differential equations 
(\ref{dirac1}, \ref{dirac2}) can again be easily solved
\begin{eqnarray}
\label{solution1}
\psi_1 &=& \left(\sqrt[4]{z \bar z}\right) 
\cons_1 ( z ) \, , \\
\label{solution2}
\psi_2 &=& \left(\sqrt[4]{z \bar z}\right)^{- 1} 
\overline{\cons_2 ( z )} \, , 
\end{eqnarray}
where $\cons_{1}$ and $\cons_2$ are analytic functions of 
$z$ and $\overline{\cons_2(z)}$ is the complex conjugate of $\cons_2(z)$. 
Choosing e.g.~$\cons_1 = 1$ we get 
$\psi_1 = \sqrt{r}$, i.e.~a real-valued function 
of $x$ and $y$. The zero modes (\ref{solution1}, \ref{solution2}) 
of the Dirac operator in the background of a single center vortex 
are not normalizable. This is because normalizable analytic functions 
$\cons_{1/2}$ have to approach zero at infinity, and 
therefore, have to be identically zero or have a pole 
which yields a non-integrable singularity as was already discussed
above. 

This result is in accord with the index theorem for $U(1)$ gauge fields 
in $D=2$. The theorem states that the difference between the
numbers of right- and left-handed fermionic zero modes, say $n$, 
in the background
of a $U(1)$ gauge potential on $\R^2$ is related to the total flux
$\Phi = \frac{1}{2 \pi i} \int_{\R^2} F$ by \cite{index-theorem-D=2}
\begin{equation}
n = \left[ \Phi \right] \, , 
\end{equation}
where $[x]$ is the largest integer smaller than $x \in \R$ (i.e.~$[1] =
0$). 
In the case of a single center vortex on $\R^2$ the total flux is 
equal to $1/2$, i.e.~the number of left-handed zero modes is equal to
the number of right-handed zero modes (in the present case there are no
normalizable left- or right-handed zero modes).
There exists an extensive literature about the Dirac operator in the 
background of Abelian gauge fields in two dimensions. A few remarks
about some papers concerning also eigenmodes of such Dirac operators 
are in order here. In \cite{Ma-86} the authors 
considered zero modes of the Dirac operator in the
background of a flux tube on a disc with finite radius and APS boundary
conditions. In this case the index $n$ of the Dirac operator receives
contributions from the boundary of the space-time manifold\footnote{We
consider in the present paper space-time manifolds with the topology 
$\R^N$, $S^N$ or $\T^N$. The latter two have no boundary whereas on 
$\R^N$ the gauge potential
vanishes at infinity. Therefore in our case there are no boundary
contributions to the index.}. In
\cite{Sitenko-96} self-adjoint extensions of the
Dirac operator in the background of a singular magnetic vortex have been
examined. Due to the boundary conditions used in these papers the 
encountered eigenmodes differ from the ones found here. 
In our opinion it is not
clear which boundary conditions are the correct ones in a physical
context.
Therefore we will consider a smooth gauge potential 
representing a smeared out vortex. This way we get rid of the 
singularity of the vortex field and avoid the discussion of 
self-adjoint extensions of the Dirac operator at the position 
of the vortex.
We choose a profile function 
$f(r^2) = r^2 / ( \varepsilon^2 + r^2 ) \, , \, \varepsilon \in \R$,
i.e.~we work with the gauge potential
\begin{equation}
\label{vortex2}
A_x = - \frac{y}{r^2 + \varepsilon^2} \frac{i}{2} \, , \quad 
A_y = \frac{x}{r^2 + \varepsilon^2} \frac{i}{2} \, .
\end{equation}
For this profile function $f$ we obtain 
$\phi(r^2) = \ha \log (r^2 + \varepsilon^2 )$.
For $\lambda=0$ the differential equations 
(\ref{dirac1}, \ref{dirac2}) can be solved
also in this case:
\begin{eqnarray}
\label{2.16}
\psi_1 &=& \left( \sqrt[4]{r^2+\varepsilon^2} \right)
\tilde \cons_1 ( z ) \, ,\\
\label{2.17}
\psi_2 &=& \left( \sqrt[4]{r^2+\varepsilon^2} \right)^{- 1} 
\overline{\tilde \cons_2 ( z )} \, .
\end{eqnarray}
For $\tilde \cons_{1/2} = 1$ these solutions are obviously not
normalizable. With the same arguments as before we conclude that 
there are no normalizable zero modes.

As discussed already above in connection with the index theorem 
the asymptotic behavior of the zero mode of the Dirac operator changes
if we change the magnitude of the magnetic flux. 
Multiplying the gauge potential $A$ 
in (\ref{vortex2}) by a factor $\rho \in \R^+$ 
the flux $\Phi$ becomes $\rho / 2$ and 
the solutions $\psi_{1/2}$, cf.~eqs.~(\ref{2.16}, \ref{2.17}),
change to $(\psi_{1/2})^\rho$. This means that for $\rho > 2$ 
we get a total flux $\Phi > 1$ and 
a normalizable zero mode with chirality $-1$: 
\begin{eqnarray}
\psi_2 &=& (\sqrt[4]{r^2+\varepsilon^2})^{- \rho} \, , \quad 
\psi_1 \equiv 0 \, .
\end{eqnarray}
The probability density of this zero mode has a maximum at $r=0$,
i.e.~at the location of the vortex.

Since the differential equations (\ref{dirac1}, \ref{dirac2}) 
are of first order type it is also possible to explicitly 
write down zero
modes for multi-vortex background fields, because the multi-vortex gauge
potential can be written as the sum over the gauge potentials of several
simple vortices. The solution of the corresponding multi-vortex 
differential equation is then simply the product of the solutions to the
several one-vortex differential equations. As an example take two
vortices, one at $z= a$ and a second at $z = b$. 
The solutions $\psi_{1/2}$ then read
\begin{eqnarray}
\label{zero-1}
\psi_1 &=& 
\left( \sqrt[4]{(z-a)(\bar z-\bar a) + \varepsilon^2} 
\sqrt[4]{(z-b)(\bar z-\bar b) + \varepsilon^2} \right) 
\cons_1 (z)
\, , \\
\label{zero-2}
\psi_2 &=& \left( \sqrt[4]{(z-a)(\bar z-\bar a) + \varepsilon^2} 
\sqrt[4]{(z-b)(\bar z-\bar b) + \varepsilon^2} \right)^{- 1} 
\overline{\cons_2(z)}
\, .
\end{eqnarray}

\begin{figure}
\begin{minipage}{7cm}
\centerline{\epsfxsize=7 cm\epsffile{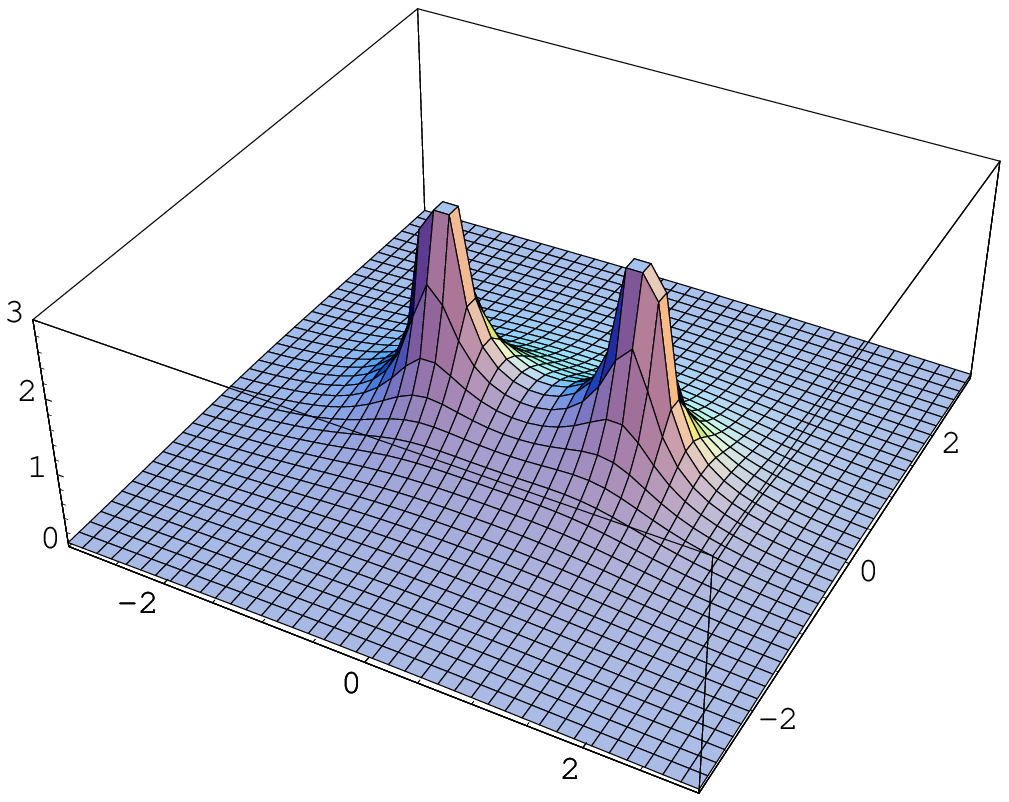}}
\caption{\label{2-vortex-prob}\textsl{Probability density of the zero
mode in the background of two center vortices in $D=2$ (see
eqs.~(\ref{zero-1}, \ref{zero-2})) for 
$\varepsilon=0.01$, $a=1$ and $b=-1$.}}
\end{minipage}
\hfill
\begin{minipage}{7cm}
\centerline{\epsfxsize=7 cm\epsffile{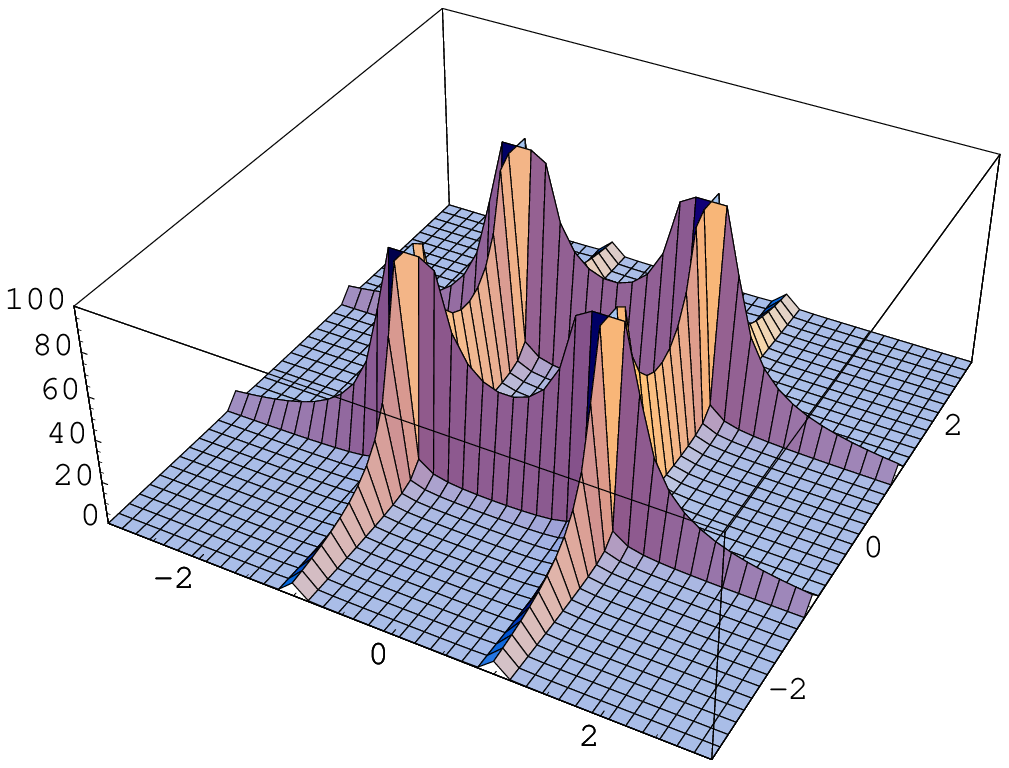}}
\caption{\label{vortex-intersection}\textsl{%
Probability density of the zero mode in the background of four 
intersecting vortex sheets is 
shown in the subspace $x_4=x_2=0$ for 
$a=b=1$ and $\varepsilon=0.01$.}}
\end{minipage}
\end{figure}

There is one quasi-normalizable\footnote{Strictly speaking we have to
increase the magnetic flux infinitesimally to get a normalizable zero
mode.} zero mode: $(0,\psi_2)$. This zero mode is obviously localized 
at the centers of the vortices (at $z=a$ and $z=b$) and on the line 
between them, cf.~fig.~\ref{2-vortex-prob}.
A similar calculation on the compact sphere $S^2$ yields a normalizable 
zero mode, cf.~Appendix \ref{zero-S2}.
According to the index theorem we obtain normalizable zero modes as soon
as the net flux of the vortices exceeds 1 (i.e.~two units of the flux of a
center vortex). 

\section{Fermionic zero modes of the Dirac operator for 
intersecting center vortex fields}
\label{modes-d4}

So far we have considered the fermionic modes in the background of
parallel non-intersecting flat center vortex sheets. Due to the 
translational
invariance parallel to the vortex sheets it was sufficient to study the
Dirac operator in the plane perpendicular to the vortex sheet, where the
vortices appear as intersection points. In the confined phase the
center vortices percolate \cite{Langfeld}. 
The vortices then have arbitrary directions and also intersect. 
In intersection points 
all four space directions participate and, consequently to study the 
fermionic modes in the background of intersecting vortices we have 
to consider the full 4-dimensional Dirac operator.

In the following we consider four vortex sheets consisting of two
orthogonal pairs of parallel vortex sheets. One vortex pair is
given by two planes parallel to the $x_1-x_2$ plane located at $x_4 = 0$
and $x_3 = \pm a$. The other vortex pair consists of two planes parallel
to the $x_3-x_4$ plane located at $x_2 = 0$ and $x_1 = \pm b$. The four
vortices intersect in four points given by $x_4 = x_2 = 0$, 
$ x_3 = \pm a$, $x_1 = \pm b$. Each of the intersection points carries
topological charge $\pm 1/2$. If we choose the orientation of the flux
of the two vortex sheets to be equal we find a total
Pontryagin index of $\nu = \pm 2$, otherwise (if the fluxes are 
anti-parallelly oriented) the Pontryagin index vanishes. For definiteness 
we choose the direction of the flux in
parallel vortex pairs to be the equal. Then the gauge potential can be
chosen as
\begin{eqnarray}
\label{multi-vortex}
&\begin{array}{rclrcl}
A_1 &= \displaystyle{\left[ - x_2  \frac{f(s_+^2)}{s_+^2}
                - x_2  \frac{f(s_-^2)}{s_-^2} \right] \frac{i}{2} }
\, ,  \quad 
A_2 =& \displaystyle{\left[ (x_1+b)  \frac{f(s_+^2)}{s_+^2} +
               (x_1-b)  \frac{f(s_-^2)}{s_-^2} \right] \frac{i}{2} }
\, , \\
\\
\nonumber
A_3 &= \displaystyle{\left[ - x_4  \frac{f(r_+^2)}{r_+^2}
                - x_4  \frac{f(r_-^2)}{r_-^2} \right] \frac{i}{2} }
\, ,  \quad 
A_4 =& \displaystyle{\left[ (x_3+a)  \frac{f(r_+^2)}{r_+^2} +
               (x_3-a)  \frac{f(r_-^2)}{r_-^2} \right] \frac{i}{2} }
\, , 
&\end{array}
\end{eqnarray}
where 
\begin{eqnarray}
r_\pm^2 &=& (x_3 \pm a)^2 + x_4^2 \, , \quad
s_\pm^2 = (x_1 \pm b)^2 + x_2^2 \, , \quad
a , b \in \R^+ \, , \\
f(r^2) &=& r^2 / (r^2 + \varepsilon^2) \, .
\end{eqnarray}
The field strength $F_{\mu \nu}$ of the above gauge potential is given by
\begin{eqnarray}
\nonumber
F_{1 2} &=& \frac{i}{2}
\left( \frac{2 \varepsilon^2}{(s_+^2 + \varepsilon^2)^2} + 
\frac{2 \varepsilon^2}{(s_-^2 + \varepsilon^2)^2} \right)
\\
\nonumber 
F_{3 4} &=& \frac{i}{2}
\left( \frac{2 \varepsilon^2}{(r_+^2 + \varepsilon^2)^2} + 
\frac{2 \varepsilon^2}{(r_-^2 + \varepsilon^2)^2} \right)
\, .
\end{eqnarray}
Splitting the field strength into (anti-)self-dual components 
\begin{equation}
F_{\mu \nu} = \left( 
\alpha \eta^3_{\mu \nu} + \beta \bar \eta^3_{\mu \nu}
\right) \, ,
\end{equation}
where $\eta^a_{\mu \nu}$ and $\bar \eta^a_{\mu \nu}$ are the t'~Hooft
symbols and $\alpha = F_{1 2} + F_{3 4} $, 
$\beta = F_{1 2} - F_{3 4}$, one finds that in the limit 
$\varepsilon \to 0$ the field strength is self-dual ($\beta = 0$) at the 
intersection points.

To determine the zero modes of the Dirac operator in the background of 
the considered intersecting vortices we introduce complex variables 
$\u$ and $\v$ \cite{Wipf}
\begin{equation}
\u = x_3 + i x_4 \, , \quad \v = x_1 + i x_2 \, 
\end{equation}
and the corresponding complex derivatives $\partial_{\u}$ and 
$\partial_{\v}$
\begin{equation}
\label{complex-u-v}
\partial_{\u} = \frac{1}{2} \left(
\partial_{x_3} - i \partial_{x_4} \right) 
\, , \quad 
\partial_{\v} = \frac{1}{2} \left(
\partial_{x_1} - i \partial_{x_2} \right) \, .
\end{equation}

We also introduce two complexified gauge potentials $A_\u$ and $A_\v$ as
in equation (\ref{complexified}). These gauge potentials can be written as
\begin{eqnarray}
A_\u & = \frac{1}{2} \partial_\u \phi \, , \quad  
A_\v & = \frac{1}{2} \partial_\v \phi \, ,
\end{eqnarray}
where 
\begin{eqnarray}
\nonumber
\phi = \phi ( \u, \bar \u, \v, \bar \v ) &= 
\log
\Big[ &
((\u - a)(\bar \u -a) +\varepsilon^2 )%
((\u + a)(\bar \u +a) +\varepsilon^2 ) 
\\
&&
((\v - b)(\bar \u -b) +\varepsilon^2 )%
((\u + b)(\bar \u +b) +\varepsilon^2 ) \Big] \, .
\end{eqnarray}

The Dirac operator in the background of the gauge potential 
(\ref{multi-vortex}) reads
\begin{eqnarray}
i \gamma_\mu \D_\mu \psi &=&
i \left( \begin{array}{cccc}
        0 & 0 & i \D_4 + \D_3 & \D_1 - i \D_2 \\
        0 & 0 & \D_1 + i \D_2 & i \D_4 - \D_3 \\
        -i \D_4 + \D_3 & \D_1 - i \D_2 & 0 & 0 \\
        \D_1 + i \D_2 & - i \D_4 - \D_3 & 0 & 0 \\
        \end{array} 
\right) \,
\left( \begin{array}{c}
        \psi_1 \\
        \psi_2 \\
        \psi_3 \\
        \psi_4
        \end{array}
\right)  
\, ,
\end{eqnarray}
where $\D_\mu = \partial_\mu + A_\mu$ is the covariant derivative.
For the considered vortex gauge potential (\ref{multi-vortex}) the
eigenvalue equation (\ref{Dirac-equation}) for the Dirac operator 
reduces to four coupled differential equations:
\begin{eqnarray}
\nonumber
\lambda \psi_1 &=& 
i \left( 2 \partial_{\bar \u} -
\frac{1}{2} \partial_{\bar \u} \phi
\right) \psi_3 + 
i \left( 2 \partial_{\v} +
\frac{1}{2} \partial_{\v} \phi
\right) \psi_4 \, ,
\\
\nonumber
\lambda \psi_2 &=&
i \left( 2 \partial_{\bar \v} -
\frac{1}{2} \partial_{\bar \v} \phi 
\right) \psi_3 + 
i \left( - 2 \partial_{\u} -
\frac{1}{2} \partial_{\u} \phi 
\right) \psi_4 \, , 
\\
\nonumber
\lambda \psi_3 &=& 
i \left( 2 \partial_{\u} +
\frac{1}{2} \partial_{\u} \phi 
\right) \psi_1 + 
i \left( 2 \partial_{\v} +
\frac{1}{2} \partial_{\v} \phi 
\right) \psi_2 \, , 
\\
\nonumber
\lambda \psi_4 &=&
i \left( 2 \partial_{\bar \v} -
\frac{1}{2} \partial_{\bar \v} \phi
\right) \psi_1 + 
i \left( - 2 \partial_{\bar \u} + 
\frac{1}{2} \partial_{\bar \u} \phi
\right) \psi_2 \, ,
\end{eqnarray} 
which can be easily solved for $\lambda = 0$, since in this case the
upper and lower components of the Dirac spinor decouple. 
The $\lambda = 0$ solutions read
\begin{eqnarray}
\nonumber
\psi_1 &=& 
\left(
{\sqrt[4]{(r_+^2+\varepsilon^2)(r_-^2+\varepsilon^2)}}
\right)^{- 1} 
{\sqrt[4]{(s_+^2+\varepsilon^2)(s_-^2+\varepsilon^2)}} 
\cons_1 ( \bar \u , \v ) 
\, , \\
\nonumber
\psi_2 &=& 
{\sqrt[4]{(r_+^2+\varepsilon^2)(r_-^2+\varepsilon^2)}} 
\left(
{\sqrt[4]{(s_+^2+\varepsilon^2)(s_-^2+\varepsilon^2)}}
\right)^{- 1} 
\cons_2 ( \u , \bar \v ) 
\, , \\
\nonumber
\psi_3 &=& 
{\sqrt[4]{(r_+^2+\varepsilon^2)(r_-^2+\varepsilon^2)}} 
{\sqrt[4]{(s_+^2+\varepsilon^2)(s_-^2+\varepsilon^2)}} 
\cons_3 ( \u , \v ) 
\, , \\
\nonumber
\psi_4 &=& 
\left(
{\sqrt[4]{(r_+^2+\varepsilon^2)(r_-^2+\varepsilon^2)}}
\right)^{- 1} 
\left(
{\sqrt[4]{(s_+^2+\varepsilon^2)(s_-^2+\varepsilon^2)}}
\right)^{- 1} 
\cons_4 ( \bar \u , \bar \v ) 
\, .
\end{eqnarray}
The analytic functions 
$\cons_i$ have to be chosen constant to avoid non-integrable
singularities as was discussed for the $D=2$ case. 
The only spinor component going to zero at infinity is 
$\psi_4$. Therefore, only the spinor with
non-vanishing component $\psi_4$ 
(and all other components zero) 
yields  a normalizable zero mode (normalizable up
to a logarithmic divergence from the integration $r_\pm \to \infty$ and 
$s_\pm \to \infty$ - but these divergences become integrable if the
vortex gauge potential is multiplied with $\rho = 1 + \alpha \, ,
\, \alpha > 0$). 
Embedding the $U(1)$ gauge group into an $SU(2)$ gauge group 
results in  a second zero 
mode with opposite iso-spin (corresponding to inverting the sign in
front of the vortex gauge potential (\ref{multi-vortex})) and the same
chirality as the above zero mode. This second zero mode is given 
by\footnote{Embedding the $U(1)$ gauge group into $SU(2)$ gives the
Pontryagin index 2 for this configuration. This corresponds to the 
$2$ observed zero modes.}
\begin{eqnarray}
\nonumber
\psi_1 &=&  \psi_2 = \psi_4 \equiv 0
\, , \\
\nonumber
\psi_3 &=& {\left(\sqrt[4]{(r_+^2+\varepsilon^2)%
(r_-^2+\varepsilon^2)}\right)}^{-1} 
{\left(\sqrt[4]{(s_+^2+\varepsilon^2)(s_-^2+\varepsilon^2)}
\right)}^{-1} 
\, .
\end{eqnarray}

The probability density of the zero modes is peaked at the four 
vortex intersection points and at the vortex sheets, 
cf.~fig.~\ref{vortex-intersection}.

\section{Center vortices and Dirac equation on the 2-torus}
\label{modes-t2}

In this chapter we consider zero modes
of the Dirac operator in the background of Abelian gauge potentials 
representing Dirac strings and center vortices on the torus $\T^2$. 
There is a variety of reasons for studying $\T^2$ in addition to
$\R^2$: first, $\T^2$ allows to use the Atiyah-Singer index theorem in
a stringent fashion and allows for normalizable spinors in the
background of integer flux. Second, from a physical point of view one
would want the mechanism of chiral symmetry breaking to depend on
local quantities rather than on global characteristics like boundary
conditions imposed from the manifold.  
However, individual solutions
of differential equations may depend strongly on the boundary
conditions. So if the envisaged mechanism of chiral symmetry breaking
contains a truncation to a subset of modes of the Dirac operator, it
is not a priori clear that the mechanism is indeed independent of 
the boundary conditions.
Therefore the proposed mechanism
has to be checked for different topologies of the space-time manifold.
Third, the torus simulates a periodic arrangement of vortices. This is
much closer to a percolated vortex cluster than a single vortex in
$\R^2$. It will turn out that the zero modes are again localized at
the position of the vortex, thus strengthening the point of view that
the zero modes are indeed influenced by local properties of the gauge
potential only. Finally, the torus is the space-time
manifold that is also used in lattice calculations. 

At this point it should be mentioned that zero modes of the Dirac
operator in the background of ideal vortices on the torus are also 
relevant to find instantons on a (dual) torus by the Nahm
transformation \cite{Ford-00,Ford-02}.

\subsection{Periodicity properties of the gauge potential on the torus}

We first consider flat vortices so that we can restrict ourselves to a
two-torus where the vortices appear as piercing points. 
Instead of working on the compact torus we can work on its 
covering manifold - on $\tilde \T^2 = \R^2 \equiv \C$. 
But in this case we have to demand that 
physical, i.e.~gauge invariant, observables are periodic. 
This implies that the gauge potential has to be periodic up to 
gauge transformations.  We choose the two-torus to have circumferences 
$1$ and $\tau$. Then the gauge potential $A$ has to fulfill 
``quasi-periodicity'' conditions
\begin{eqnarray}
\label{period-a}
A(z+1) = A^{U_x(z)} (z) \, , \quad  A(z+i \tau) = A^{U_y(z)} (z) \, ,
\end{eqnarray}
where $A_\mu^U=U^{-1} A_\mu U + U^{-1} \partial_\mu U$ is the gauge 
transform of $A$. The transition functions $U_x , U_y$ have to fulfill the 
cocycle condition \cite{vBaal}
\begin{eqnarray}
U_x (z) U_y (z+1) = U_y (z) U_x (z+i \tau) \, .
\end{eqnarray}
We can always choose $U_x = \Id$ and $U_y$ independent of $y$. 
This implies periodicity of $A(z)$ in $x$-direction. Furthermore,  
$U_y$ as function of $x$ defines a mapping from $S^1$ into $U(1)$. 
Such mappings fall into homotopy classes $\pi_1(U(1)) = \pi_1 (S^1)$
which are characterized by a winding number $n$. 
A simple calculation shows that this winding number is related to the
magnetic flux $\Phi$ through the torus
\begin{equation}
\label{flux-1}
\Phi = - n \, ,
\end{equation}
reflecting the quantization of the magnetic flux through the torus.

\subsection{Dirac string on the torus}

Before we
write down the gauge potential of center vortices on the torus let us
consider the gauge potential of a single Dirac string. The reason is
that a single center vortex does not exist on a torus while a single
Dirac string does. 

On the two-torus the gauge potential of a singular point-like object can
be expressed by means of the theta functions
\begin{equation}
\theta ( z , i \tau ) = \sum_{n \in \Z} 
e^{- \pi \tau n^2 + 2 \pi i n z} \, , \quad \tau \in \R_+
\end{equation}
which are analytic in $z$ and obey the periodicity properties
\begin{equation}
\label{period-theta}
\theta ( z + 1 , i \tau ) = \theta ( z , i \tau ) \, , \quad 
\theta ( z + i \tau , i \tau ) = e^{\pi \tau - 2 \pi i z}
\theta ( z , i \tau ) \, .
\end{equation}
The only zeros of this function are at the points \cite{tata}
\begin{equation}
z = (m + 1/2) + (n + 1/2) i \tau \, , \quad m,n \in \Z \, .
\end{equation}
For subsequent consideration let us also introduce the real-valued
function \cite{Ford-00,Ford-02}
\begin{equation}
\phi_0 (z,z_0) = \frac{1}{2} 
\log{\left( \theta_\tau (z,z_0) 
\overline{\theta_\tau(z,z_0)} \right)} \, , \quad 
\theta_\tau (z,z_0) := \theta(z + 1/2 + 1/2 i \tau - z_0, i \tau)
\, ,
\end{equation}
which (up to a factor of $2\pi$) represents a (non-periodic) Greens
function of the Laplacian on the two-torus\footnote{On the torus there 
is no periodic Greens function of the Laplacian}. By Taylor expanding 
$\theta_\tau (z,z_0) $ around its zero at $z=z_0$ one
finds that the Greens function $\phi_0 (z,z_0)$ behaves near $z_0$ as
$\log|z-z_0|$. By means of this Greens function the gauge potential of a
Dirac string on the two-torus can be expressed as 
\begin{equation}
\label{gauge-potential1}
A_x = - i \partial_y \phi_0 \, , \quad A_y = i \partial_x \phi_0 \, .  
\end{equation} 
Using the same notation as in equation (\ref{2.2})
\begin{equation}
A_z := 1/2 ( A_x - i A_y ) \, , \quad A_x = 2 i \Im(A_z) 
\, , \quad A_y = 2 i \Re(A_z) \, ,
\end{equation}
we obtain
\begin{equation}
\label{A-z-torus}
A_z = \frac{1}{2} (- i \partial_y \phi_0 + \partial_x \phi_0 ) 
= \partial_z \phi_0 
= \frac{1}{2} 
\frac{\partial_z \theta_\tau(z,z_0)}{\theta_\tau(z,z_0)} \, ,
\end{equation}
where we used that $\theta_\tau(z,z_0)$ is an analytic function of $z$ 
and $\bar \theta_\tau(z,z_0)$ is an anti-analytic function, 
i.e.~it does not depend on $z$. The periodicity properties of 
$A_z, A_x $ and $A_y$ 
can simply be derived using equation
(\ref{period-theta})
\begin{eqnarray}
A_z ( z+1 ) &=& A_z (z)\, , \quad A_z ( z+ i \tau) =
- \pi i + A_z ( z ) \, , \\ 
A_y ( z+1 ) &=& A_y (z+i\tau) = A_y (z) \, , \\ 
A_x ( z+1 ) &=& A_x (z) \, , \quad 
A_x(z+i \tau) = - 2 \pi i + A_x (z) \, . 
\end{eqnarray}
This means that the gauge potential fulfills equation (\ref{period-a}) 
with $U_x (x,y) = \Id$ and $U_y (x,y) = \exp{(-2 \pi i x)} \Id$. 
Furthermore, computing explicitly the flux going through the torus, we obtain
(using Stokes theorem)
\begin{eqnarray}
\nonumber
\Phi &=& \frac{1}{2 \pi i} \left( 
\int_{(0,0)}^{(1,0)} A_x \d x + 
\int_{(1,0)}^{(1,\tau)} A_y \d y + 
\int_{(1,\tau)}^{(0,\tau)} A_x \d x + 
\int_{(0,\tau)}^{(0,0)} A_y \d y \right) \\
&=&
\label{flux1}
\frac{1}{2 \pi i} \left(
\int_{0}^{1} ( A_x(x,0) - A_x (x,\tau) ) \d x +
\int_{0}^{\tau} (A_y(1,y) - A_y(0,y))\d y \right) = 1 \, ,
\end{eqnarray}
which is consistent with (\ref{flux-1}). 
The Dirac string is located at the point where 
$\theta_\tau (z,z_0) = 0$, i.e.~at 
$z = z_0 + m + i n \tau \, , \, m,n \in \Z$. 
The field strength of the 
Dirac string configuration vanishes at points where 
$\theta_\tau (z,z_0) \neq 0$, because
\begin{equation}
F_{xy} = \partial_x A_y - \partial_y A_x =
i (\partial^2_x + \partial^2_y) \phi = 
\Re{(4 \partial_{\bar z} \partial_z \phi)} = 
2 \Re{\partial_{\bar z} (\partial_z \theta_\tau(z,z_0) /%
 \theta_\tau(z,z_0))} 
= 0 \, .
\end{equation}
Here we used again that $\theta(z,z_0)$ is an analytic function, 
i.e.~it is independent of $\bar z$. 
On the other hand the flux through 
the torus is $1$, cf.~(\ref{flux1}), from which we conclude 
that we have a Dirac string (which is represented by a 
point in $D=2$) at the zero of the function 
$\theta_\tau(z,z_0)$, i.e.~at $z =  z_0 + m  + n i \tau $. 

The Dirac string can also be written as a pure gauge. If we
define the $U(1)$ gauge function
\begin{equation}
g (z) = \frac{\theta_\tau (z,z_0)}{|\theta_\tau(z,z_0)|} = 
\sqrt{\frac{\theta_\tau(z,z_0)}{\overline{\theta_\tau(z,z_0)}}} 
\in U(1) \, ,
\end{equation}
which is singular at the zeros of $\theta_\tau(z,z_0)$, 
the gauge potential
\begin{equation}
A_\mu = g^{-1} \partial_\mu g \, 
\end{equation}
becomes
\begin{eqnarray}
A_z &=& \frac{1}{2} (A_x - i A_y) = g^{-1} \partial_z g =
\sqrt{\frac{\overline{\theta_\tau (z,z_0)}}{\theta_\tau (z,z_0)}}
\partial_z 
\sqrt{\frac{\theta_\tau (z,z_0)}{\overline{\theta_\tau (z,z_0)}}} 
\\
&=&
\frac{1}{2} 
\frac{\partial_z \theta_\tau (z,z_0)}{\theta_\tau (z,z_0)} \, ,
\end{eqnarray}
which agrees with equation (\ref{A-z-torus}).
The periodicity properties of $g$ are as follows:
\begin{equation}
g(z+1) = g(z) \, , \quad g(z+i \tau) = 
- e^{- \pi i ((z - z_0) + (\bar z - \bar z_0))} g(z) \, . 
\end{equation}
Since the gauge potential of the Dirac string is a
pure gauge we can simply write down the zero modes of the 
corresponding Dirac operator. The zero mode is nothing but the gauge
transformation of a constant Dirac field.

In the following we will consider only smeared out vortices on the 
torus. This way we avoid the discussion of boundary 
conditions of the the spinor field at the position of the singularity 
of an ideal vortex. This discussion would involve the problem of 
selfadjoint extension of the naive Dirac operator, see 
e.g.~\cite{Ma-86,Sitenko-96,Laenge-01}.

The gauge potential of a smeared out Dirac string at the point 
$z_0$ is given by equation (\ref{gauge-potential1}) with the function
$\phi_0(z,z_0)$ replaced by
\begin{eqnarray}
\label{phi-0}
\phi (z,z_0) &=& \frac{1}{2} 
\log{(\theta_\tau^+ (z,z_0) \overline{\theta_\tau^+(z,z_0)} + 
\theta_\tau^- (z,z_0) \overline{\theta_\tau^-(z,z_0)})} \, , \\
\label{theta-0-pm}
\theta_\tau^\pm (z,z_0) &=&  
\theta ( z + \ha + \iha \tau - z_0 \pm \varepsilon , i \tau) 
\, , \quad 
\varepsilon \in \R_+ \, ,
\end{eqnarray}
i.e.~by
\begin{eqnarray}
\label{Dirac-string}
A_z 
&=& 1/2 ( A_x - i A_y ) = 
\partial_z \phi(z,z_k) \\
&=& 
\ha \, 
\frac{\overline{\theta_\tau^+(z,z_k)} 
\partial_z \theta_\tau^+ (z,z_k) +
\overline{\theta_\tau^-(z,z_k)} 
\partial_z \theta_\tau^- (z,z_k)%
}{\theta_\tau^+ (z,z_k) 
\overline{\theta_\tau^+(z,z_k)} + 
\theta_\tau^- (z,z_k) 
\overline{\theta_\tau^-(z,z_k)}} \, .
\end{eqnarray}
By Taylor expanding around $z_k$ we obtain the behavior of 
the gauge potential for small distances $r$ from the center 
$z_0$ of the Dirac string:
\begin{equation}
A_x = - y \frac{i}{( r^2 + \varepsilon^2 )} \, , \quad 
A_y = x \frac{i}{( r^2 + \varepsilon^2 )} \, .
\end{equation}
This gauge potential indeed represents a smeared out Dirac string. 
In the limit $\varepsilon \to 0 \, $ $A$ becomes the gauge 
potential of an ideal Dirac string \cite{Engelhardt-00-1} on the torus.

\subsection{Fermionic zero modes of center vortices on the torus}

A center vortex living in the Cartan sub-algebra 
can be represented by half the gauge potential
of a Dirac string. However, for a single center vortex it is not 
possible to
relate the gauge potential at $z$ with the gauge potential at 
$z+ i \tau$
by a transition function $U_y$ which is periodic in $x$. 
Instead we need an even 
number of center vortices on the torus, in accord with the 
quantization of magnetic flux through the torus, cf.~(\ref{flux-1}).
We consider the configuration, see eqs.~(\ref{phi-0},\ref{theta-0-pm}),
\begin{eqnarray}
\label{period-vortex}
A_z &=& 
\frac{1}{2} \left( 
\partial_z \phi (z,z_1) + 
\partial_z \phi (z,z_2)
\right)  \\
\nonumber &=& 
\frac{1}{4} \, 
\sum_{k=1}^2 
\frac{\overline{\theta_\tau^+(z,z_k)} 
\partial_z \theta_\tau^+ (z,z_k) +
\overline{\theta_\tau^-(z,z_k)} 
\partial_z \theta_\tau^- (z,z_k)%
}{\theta_\tau^+ (z,z_k) \overline{\theta_\tau^+(z,z_k)} + 
\theta_\tau^- (z,z_k) \overline{\theta_\tau^-(z,z_k)}} 
\, ,
\end{eqnarray}
which consists of two (smeared out) center vortices at the points 
$z_1$ and $z_2$ and fulfills the periodicity properties
(\ref{period-a}) with transition functions $U_x = \Id$ and 
$U_y = \exp{(- 2 \pi i x)} \Id = \exp{(- \pi i (\bar z + z))} \Id$. 

We are interested in the $\lambda = 0$ eigenfunctions of the Dirac
operator in the background of these two vortices. In accord with
the boundary conditions to the gauge field (\ref{period-a}) we 
impose the boundary conditions
\begin{equation}
\label{period-psi}
\psi(z+1) = U_x(z)^{-1} \psi(z) = \psi(z) \, , \quad 
\psi(z+i \tau) = U_y(z)^{-1} \psi(z) = e^{\pi i (\bar z + z)} 
\psi(z) 
\end{equation}
to the Dirac spinor $\psi$. 
The Dirac eigenvalue equation for the two spinor components reads
\begin{eqnarray}
i ( 2 \partial_z + 2 A_z ) \psi_2 &=& \lambda \psi_1 \, , \\
i ( 2 \partial_{\bar z} + 2 A_{\bar z} ) \psi_1 &=& \lambda \psi_2 \, ,
\end{eqnarray}
where $A_{\bar z} = 1/2 ( A_x + i A_y) = - \overline{A_z}$.
Because of the simple form of the two-vortex gauge potential 
(\ref{period-vortex}) the zero modes of the Dirac operator 
can be found explicitly. A short calculation yields
\begin{eqnarray}
\label{zero-torus--}
\psi_1 &=& 
\left( \prod_{k=1}^2 \, 
\left( \theta_\tau^+ (z,z_k) \overline{\theta_\tau^+(z,z_k)} + 
\theta_\tau^- (z,z_k) \overline{\theta_\tau^-(z,z_k)} \right) 
\right)^{1/4} 
\cons_1 (z) \, ,
\\
\label{zero-torus-+}
\psi_2 &=& 
\left( \prod_{k=1}^2 \, 
\left( \theta_\tau^+ (z,z_k) \overline{\theta_\tau^+(z,z_k)} + 
\theta_\tau^- (z,z_k) \overline{\theta_\tau^-(z,z_k)} \right) 
\right)^{-1/4} 
\overline{\cons_2 (z)} \, .
\end{eqnarray}
The analytic functions $\cons_1 (z)$ and $\cons_2 (z)$ have to be
chosen such that $\psi_{1/2}$ are normalizable and fulfill the 
periodicity properties (\ref{period-psi}). To render $\psi_{1/2}$
normalizable the functions $\cons_{1/2}$ must not have 
poles\footnote{A pole of $\cons_{1/2}$ would yield a non-integrable
(logarithmically divergent) singularity in the norm of the spinor $\psi$,
because the pre-factors of $\cons_{1/2}$ in 
equations (\ref{zero-torus--},\ref{zero-torus-+}) are non-zero on the
whole torus, since we only consider smeared out vortices.}. 
Inserting equations (\ref{zero-torus-+}, \ref{zero-torus--}) 
into  (\ref{period-psi}) and 
using the periodicity properties (\ref{period-theta}) of the theta
function we arrive at
\begin{eqnarray}
\label{period-f-}
\cons_1 ( z + i \tau ) &=& 
e^{\left(- \pi \tau + 2 \pi i \left( z +  \iha \tau  - \iha \Im{(z_1 + z_2)} 
\right) \right) } \cons_1 ( z ) \, , \quad \cons_1 (z+1) = \cons_1 (z) 
\, , \\
\label{period-f+}
\cons_2 ( z + i \tau ) &=& 
e^{\left(\pi \tau - 2 \pi i \left( z + \iha \tau  - \iha \Im{(z_1 + z_2)} 
\right) \right) } \cons_2 ( z ) \, , \quad \cons_2 (z+1) = \cons_2 (z) 
\, .
\end{eqnarray}
If we require analyticity of
$\cons_2$ on the whole torus then this function is fixed (up to a factor) 
by the periodicity properties (\ref{period-f+}) to be the theta 
function 
\begin{equation}
\label{f-plus}
\cons_2 (z) = \theta(z + \iha \tau  - \iha \Im{(z_1 + z_2)},i \tau) \, .
\end{equation}
This is proven in Appendix \ref{uniqueness}. 
The function $\psi_2$, cf.~(\ref{zero-torus-+}), 
is obviously normalizable, because it has no singularities.
The only zeros of $\cons_2 (z)$ are at the points
$z = \ha + \iha \Im{(z_1 + z_2)} + m + n i \tau$. 
On the other hand the required periodicity properties for $\cons_1$ 
(\ref{period-f-})
show that there is no (non-trivial) function $\cons_1$  which is 
analytic on the whole torus. The function $\cons_1$ has to have at 
least one pole. Such a solution is given by the inverse of a theta 
function
\begin{equation}
\cons_1 (z) = 1 / \theta(z + \iha \tau  - \iha \Im{(z_1 + z_2)},i \tau) 
\, .
\end{equation}
The poles of this function are at the points
$z = z_{m n} = \ha + \iha \Im{(z_1 + z_2)} + m + n i \tau \, , \, m,n
\in \Z$. 
Our considerations show that there is only one normalizable zero mode
(with $\psi_1 \equiv 0$). 
If the component $\psi_1$ of the Dirac spinor is not identically zero 
then it would have a pole at $ z = z_{m n} $, because the pre-factor 
of $\cons_1$ in equation (\ref{zero-torus--}) is non-zero on the hole 
torus. This pole would yield a logarithmic divergence.

In Appendix \ref{modes-multi-t2} multi-vortex 
configurations and the
corresponding zero modes of the Dirac operator are presented.

There is another interesting point related to the periodicity 
properties given by (\ref{period-a}, \ref{period-psi}). Multiplying the
transition functions $U_x$ and $U_y$ by constant phases, say 
$e^{-i \alpha}$ and $e^{-i \beta}$, resp., the center vortex and
Dirac string gauge potentials, (\ref{period-vortex}) and 
(\ref{gauge-potential1}), 
still fulfill the periodicity properties with the new 
transition functions 
$\tilde U_x = e^{-i \alpha} \Id$ and 
$\tilde U_y = e^{-2 \pi i x - i \beta} \Id$.  
But these new periodicity properties change the zero modes of the Dirac
operator by changing the periodicity properties of the analytic 
functions $\cons_{1/2} (z)$. The new solution reads
\begin{equation}
\tilde \cons_2 (z) = e^{- i \alpha z} 
\theta(z + \iha \tau  - \frac{i \alpha}{2 \pi} \tau - \iha \Im{(z_1 + z_2)}
+ \frac{\beta}{2 \pi} ,i \tau)
\end{equation} 
The zeros of this function are at the points 
$z = \ha + \iha \Im{(z_1 + z_2)} + \frac{i \alpha}{2 \pi} \tau -
\frac{\beta}{2 \pi} + m + n i \tau$, i.e.~the zeros are shifted by 
$\frac{i \alpha \tau - \beta}{2 \pi}$ compared to the original case,
where $\tilde U_x = \Id$ and $\tilde U_y = e^{-2 \pi i x} \Id$.  

The change of the transition functions by multiplying with constant
phases $e^{-i \alpha}$ and $e^{-i \beta}$ 
is equivalent to introducing a constant background gauge
potential $A_z = \ha ( \frac{\beta}{\tau} + i \alpha )$ and leaving the
transition functions unchanged. 

The above considerations have shown that for a two-vortex gauge potential 
(smeared out vortices) we get exactly one normalizable zero mode which has
exactly one zero on the torus. The position of the zero depends on the
imaginary parts (y-coordinates) of the positions of the vortices and on
the periodicity properties of the gauge potential 
(or equivalently on the presence of a constant background
field) and of the spinor field. Furthermore, the
probability density of the spinor field is peaked at the positions of
the vortices.

\begin{figure}
\begin{minipage}{7cm}
\centerline{\epsfxsize=7 cm\epsffile{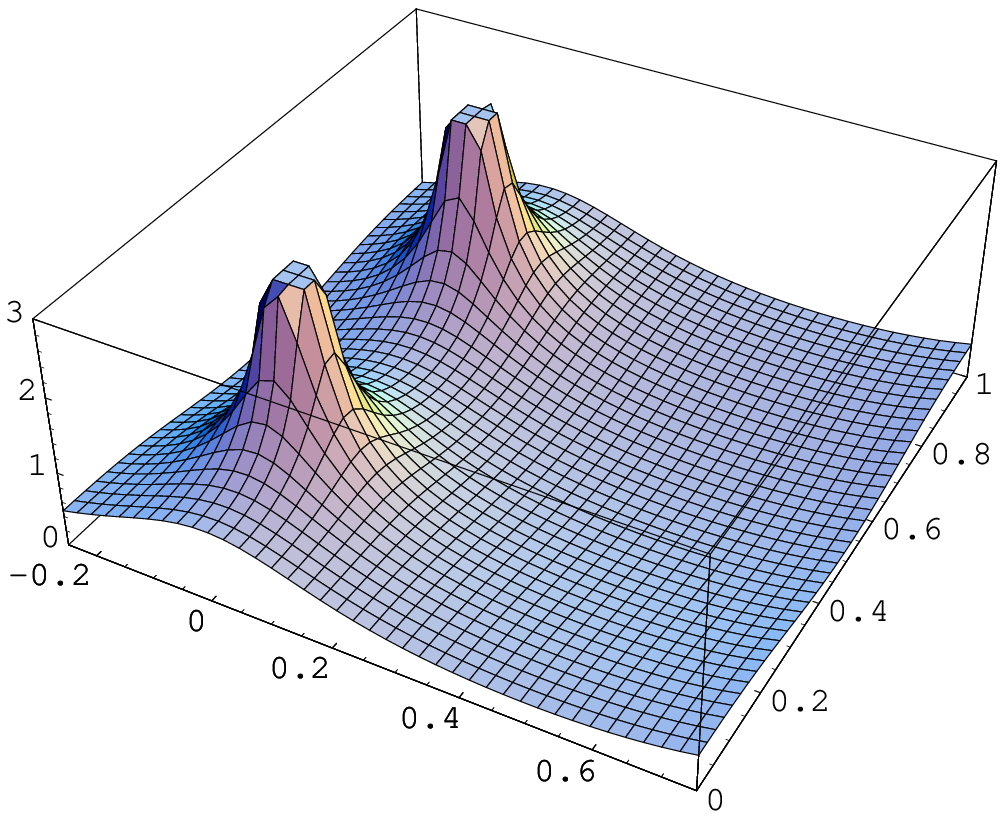}}
\caption{\label{2-vortex-prob-torus}\textsl{%
Probability density of the zero
mode in the background of two center vortices on $\T^2$ for 
$\tau = 1$, $\varepsilon = 0.01$, $z_1 = 0.25 i$ and
$z_2 = 0.75 i$.}}
\end{minipage}
\hfill
\begin{minipage}{7cm}
\centerline{\epsfxsize=7 cm\epsffile{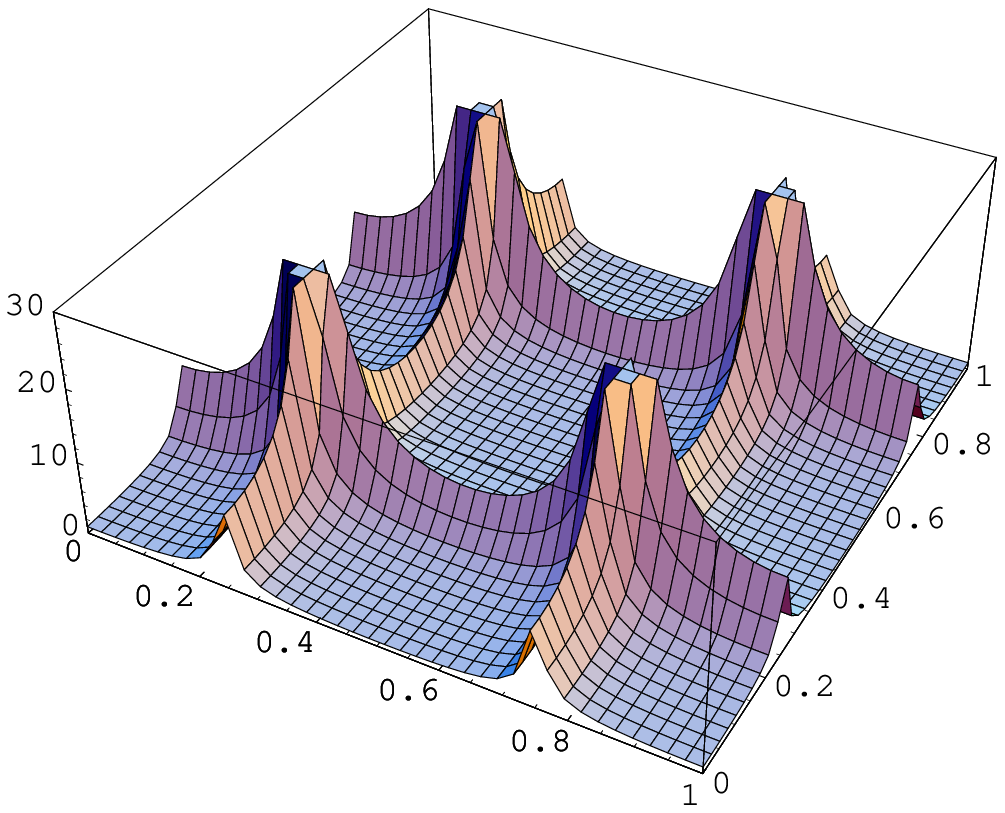}}
\caption{\label{vortex-crossing-twotorus}\textsl{%
Probability density of the zero
mode in the background of four intersecting vortex sheets  
on $\T^4$ shown in the subspace $x_1=x_3=0$ 
for $u_1=v_1=0.25 i \, , \, u_2=v_2=0.75 i$ and 
$\varepsilon = 0.01$.}}
\end{minipage}
\end{figure}

\section{Fermionic zero modes for intersecting vortices on the 4-torus}
\label{modes-t4}

As in the case of space-time $\R^4$ the fermionic zero modes for
intersecting vortices on $\T^4$ can be explicitly written down. 

We consider four smeared out center vortex sheets consisting of two
orthogonal pairs of parallel vortex sheets 
intersecting in 4 points as in section \ref{modes-d4}.
Introducing complex variables $u$ and $v$, 
cf.~eq.~(\ref{complex-u-v}), the complexified gauge potential 
$A_z$ can be chosen as 
\begin{eqnarray}
\nonumber
A_z &=& 
\partial_u ( \phi (u, u_1) + \phi(u,u_2) ) 
+ \partial_v ( \phi (v, v_1) + \phi(v,v_2) ) \, ,
\end{eqnarray}
where the function $\phi(z,z_k)$ is defined in equation (\ref{phi-0}).
As in the case of the space-time manifold $\R^4$ there is only 
one normalizable zero mode given by
\begin{eqnarray}
\nonumber
\psi_1 &=&  \psi_2 = \psi_3 \equiv 0
\, , \\
\nonumber
\psi_4 &=& 
\left( \prod_{k=1}^2 \, 
\left( \theta_\tau^+ (u,u_k) \overline{\theta_\tau^+(u,u_k)} + 
\theta_\tau^- (u,u_k) \overline{\theta_\tau^-(u,u_k)} \right) 
\right)^{-1/4} \times \\
\nonumber
&&
\times 
\left( \prod_{k=1}^2 \, 
\left( \theta_\tau^+ (v,v_k) \overline{\theta_\tau^+(v,v_k)} + 
\theta_\tau^- (v,v_k) \overline{\theta_\tau^-(v,v_k)} \right) 
\right)^{-1/4} 
\overline{\cons_4 (u,v)} \, ,
\end{eqnarray}
where $\cons_4(u,v)$ is an analytic function of $u$ and $v$ given by
\begin{equation}
\cons_4 (u,v) = 
\theta(u + \iha \tau - \iha \Im{(u_1 + u_2)},i \tau) 
\theta(v + \iha \tau - \iha \Im{(v_1 + v_2)},i \tau) 
\, .
\end{equation}
The probability distribution of this zero mode in the plane 
$\Re{u} = \Re{v} = 0$ for 
$u_1=u_2=0.25 i \, , \, u_2 = v_2 = 0.75 i$
is plotted in fig.~\ref{vortex-crossing-twotorus}.

\section{Concluding remarks}

In the present paper we have studied the properties of fermionic zero
modes in a center vortex background field. We have demonstrated that
these zero modes are concentrated at the localization of the center
vortices. In accord with this the probability density of these zero
modes is sharply peaked at the vortex intersection points which carry
(localized) topological charge $1/2$. This result is consistent with the
localization of the fermionic zero modes in an instanton background
field at the instanton center. In fact lattice calculations show a
strong correlation between the topological charge density distribution
and the distribution of the quark condensate $\langle \bar q(x) q(x) \rangle$, 
the order
parameter of chiral symmetry breaking. Given the localization of the
quark zero modes at the localization of topological charge, we expect
the quark zero modes in the vortex background field to play a crucial
role for the spontaneous breaking of chiral symmetry in the vortex
picture. This will be subject to future investigations.  

\section{Acknowledgments}

One of the authors (H.R.) is grateful to I.~Zahed for stimulating
discussions as well as for the hospitality extended to him during a 
visit at Stony Brook. V.~Ch.~Zh. gratefully acknowledges the hospitality,
extended to him by the theory group of the Institut f\"ur Theoretische
Physik, Universit\"at T\"ubingen during his stay there. 
The authors are also grateful to C.~Ford and J.M.~Pawlowski for drawing
our attention to reference  \cite{Ford-02} and to the existence
of additional zero modes in the case of ideal (singular) center
vortices. This work has
been supported by the Deutsche Forschungsgemeinschaft under grants 
DFG-Re 856/4-1 and DFG 436 RUS 113/477/4.

\appendix

\section{Conventions}
\label{conventions}

We choose the generators of the gauge group to be anti-hermitian.
Therefore the components $A_\mu$ of the gauge potential are
anti-hermitian, e.g.~purely imaginary for the gauge group $U(1)$. The
magnetic flux $\Phi$ through a closed loop ${\mathcal C}$ is 
defined by 
\begin{equation}
\label{flux-def}
\Phi = \frac{1}{2 \pi i} \oint_{\mathcal C} A_\mu \d x_\mu 
\end{equation}
and thus real valued.

The complex conjugate of the complex number z is denoted by $\bar z$.
Furthermore, $\Re(z)$ and $\Im(z)$ denote real and the imaginary part of
$z$, respectively.

We consider the Dirac equation in Euclidean space-time. In $D=2$ we
choose the $2\times2$ Dirac matrices
\begin{equation}
\label{dirac-matrices-D2}
\gamma_1 = \left( \begin{array}{cc}
                        0 & 1 \\
                        1 & 0 
                   \end{array}
           \right) \, , \quad
\gamma_2 = \left( \begin{array}{cc}
                        0 & -i \\
                        i & 0 
                   \end{array} 
	   \right) \, , \quad
\gamma_5 = - i \gamma_1 \gamma_2 
	  = \left( \begin{array}{cc}
                        1 & 0 \\
                        0 & -1 
                   \end{array} 
           \right) \, .
\end{equation}
In $D=4$ we use the chiral representation for the Dirac matrices:
\begin{eqnarray}
\label{dirac-matrices-D4}
\gamma_4 &=& \left( \begin{array}{cc}
                0 & i \Id \\
                - i \Id & 0 
                \end{array} \right) \, , \quad 
\gamma_i = \left( \begin{array}{cc}
                0 & \sigma_i \\
                \sigma_i & 0 
                \end{array} \right) \, , \quad 
i = 1,2,3 \, , \quad
\gamma_5 = \left( \begin{array}{cc}
                \Id & 0 \\
                0 & - \Id 
                \end{array} \right) \, , 
\end{eqnarray}
where $\Id$ is the $2 \times 2$ unit matrix and $\sigma_i$ are the
(hermitian) Pauli matrices.

\section{Dirac zero modes on $S^2$}
\label{zero-S2}

We consider the Dirac operator on the sphere $S^2$ with radius $R$. 
We use stereographic coordinates $x_{1/2}$ on $S^2$ defined by
\begin{equation}
\label{stereographic-coordinates}
y_i = \frac{2 R^2}{R^2 + x^2} x_i \, , \quad i=1,2 \, , \quad
y_3 = \frac{R^2-x^2}{R^2+x^2} R \, , \quad x^2 = x_1^2 + x_2^2 
\, ,
\end{equation}
where $\vec y = ( y_1,y_2,y_3)$ 
is the vector in $\R^3$ of the corresponding point of the embedded
sphere $S^2$ with radius $R$. The metric on $S^2$ in stereographic
coordinates has the form 
\begin{equation}
\d s^2 =   \Omega_R^2 ( \d x_1^2 + \d x_2^2 ) \, ,
\end{equation}
where $\Omega_R = 2 R^2 / (R^2+x^2)$. The Dirac operator 
$\hat \D$ in these coordinates is given by \cite{Abdalla}
\begin{equation}
\hat \D = \Omega_R^{-3/2} \D \Omega_R^{1/2} \, ,
\end{equation}
where $\D$ is the Dirac operator on $\R^2$. Therefore the 
zero modes $\hat \psi$ of the Dirac operator on $S^2$ are related to the
zero modes $\psi$ of the Dirac operator on $\R^2$ by
\begin{equation}
\hat \psi = \Omega_R^{-1/2} \psi \, .
\end{equation}
With $z = x_1 + i x_2$ the zero mode of the Dirac operator in the 
presence of two smeared out center vortices at the points $z=a$ 
and $z=b$ reads  
\begin{equation}
\psi_1 \equiv 0 \, , \quad 
\psi_2 = \frac{\sqrt{z \bar z + R^2}}%
{\sqrt[4]{(z-a)(\bar z-\bar a) + \varepsilon^2} 
\sqrt[4]{(z-b)(\bar z-\bar b) + \varepsilon^2}} 
\, .
\end{equation}
This zero mode is normalizable with respect to the measure on $S^2$
which is given by $\Omega_R^2 \d x_1 \d x_2$.

\section{Uniqueness of $\cons_2$ from periodicity properties}
\label{uniqueness}

The existence of an analytic function
with the periodicity properties (\ref{period-f+}) is seen by choosing 
\begin{equation}
\label{f-plus-a}
\cons_2 (z) = \theta(z + \iha \tau - \iha \Im{(z_1 + z_2)},i \tau) \, ,
\end{equation}
cf.~eq.~(\ref{period-theta}). 
To show the uniqueness (up to a constant factor) of $\cons_2$
we assume that there is another analytic function $\tilde \cons_2$ 
satisfying equation (\ref{period-f+}). Now consider the meromorphic 
function $ f(z) := \tilde \cons_2 (z) / \cons_2 (z) $. This is 
an elliptic function with periods $1$ and $i \tau$. But $f(z)$ has only
a single pole in the fundamental domain ($0 < Re(z) < 1 \, $ , $\,
0 < \Im(z) < \tau$) at the zero of
$ \theta ( z + \iha \tau - \iha \Im{(z_1 + z_2)}, i \tau) $.
But an elliptic function has to have at least two poles in the
fundamental domain or it is a constant \cite{Hurwitz}. 
Hence, we infer that 
$f(z)$ is a constant and $\cons_2 (z)$ is (up to a constant factor) 
given by (\ref{f-plus-a}).

\section{Multi vortex solution on $\T^2$}
\label{modes-multi-t2}

Let us assume we have a number of thick vortices and anti-vortices 
at the points $z_k \, , \, k=1, \ldots , m^+ $ and 
$z_l \, , \, l = m^+ +1 , \ldots , m^+ + m^-$, respectively, 
where the total number $m^+ + m^- = m $ of vortices is even. 
The corresponding gauge potential reads
\begin{equation}
A_z = \frac{1}{4} \partial_z \left( 
\sum_{k=1}^{m^+}  \phi(z,z_k) - 
\sum_{l=m^+ + 1}^{m^+ + m^-} \phi (z,z_l) \right) \, ,
\end{equation}
where $\phi(z,z_k)$ is defined by equation (\ref{phi-0}).
The fermionic zero modes $\psi_{1/2}(z)$ are given by
\begin{eqnarray}
\nonumber
\psi_1 &=& 
\prod_{k=1}^{m^+} 
\left( \theta_\tau^+ (z,z_k) \overline{\theta_\tau^+(z,z_k)} + 
\theta_\tau^- (z,z_k) \overline{\theta_\tau^-(z,z_k)}) \right)^{1/4} 
\times \\
\nonumber
&& \times
\prod_{l=m^+ + 1}^{m^+ + m^-}
\left( \theta_\tau^+ (z,z_l) \overline{\theta_\tau^+(z,z_l)} + 
\theta_\tau^- (z,z_l) \overline{\theta_\tau^-(z,z_l)}) \right)^{-1/4} 
\cons_1 (z) \, , \\
\nonumber
\psi_2 &=& 
\prod_{k=1}^{m^+} 
\left( \theta_\tau^+ (z,z_k) \overline{\theta_\tau^+(z,z_k)} + 
\theta_\tau^- (z,z_k) \overline{\theta_\tau^-(z,z_k)}) \right)^{-1/4} 
\times \\
\nonumber
&& \times
\prod_{l=m^++1}^{m^+ + m^-}
\left( \theta_\tau^+ (z,z_l) \overline{\theta_\tau^+(z,z_l)} + 
\theta_\tau^- (z,z_l) \overline{\theta_\tau^-(z,z_l)}) \right)^{1/4} 
\overline{\cons_2 (z)} \, , 
\end{eqnarray}
where $\theta^\pm$ is defined by equation (\ref{theta-0-pm}).
The periodicity properties of $\psi_{1/2}$, cf.~eq.~(\ref{period-psi}), 
define the periodicity
properties of the analytic functions $\cons_{1/2} (z)$:
\begin{eqnarray}
\cons_1 ( z + i \tau ) &=& 
e^{- \left((\Delta m / 2) ( \pi \tau - 2 \pi i ( z + i \tau /2 ) ) 
- \pi \Im{\left( \sum_{k=1}^{m^+} z_k - \sum_{l=m^+ + 1}^{m} 
z_l \right) } \right) } \cons_1 (z) \, , \\
\cons_1 (z+1) &=& \cons_1 (z) \, , \\
\cons_2 (z+i \tau) &=&
e^{\left((\Delta m / 2) ( \pi \tau - 2 \pi i ( z + i \tau /2 ) ) 
- \pi \Im{\left(\sum_{k=1}^{m^+} z_k - \sum_{l=m^+ + 1}^{m}
z_l \right)} 
\right) } \cons_2 ( z ) \, , \\
\cons_2 (z + 1) &=& \cons_2 (z) \, , 
\end{eqnarray}
where $\Delta m = m^+ - m^-$ and $m = m^+ + m^-$.
If $\Delta m >0$ we find a $\Delta m /2$-dimensional 
vector space of left-handed zero modes $\psi_2$ with analytic 
functions $\cons_2 (z)$. 
In the case 
$\Delta m<0$ we find a $\Delta m /2$-dimensional 
vector space of right-handed zero modes with analytic functions 
$\cons_1 (z)$. The functions 
$\cons_{1/2}$ are then given by products of theta functions.
In the case $\Delta m >0$ we obtain
\begin{equation}
\label{cons-multi-vortex-1}
\cons_2 (z) = \prod_{j=1}^{\Delta m/2} 
\theta ( z + i \tau /2 - \tilde z_j , i \tau) \, ,
\end{equation}
where the complex numbers $\tilde z_j$ have to fulfill the conditions
\begin{equation}
\label{cons-multi-vortex-2}
\Re{(\sum_{j=1}^{\Delta m/2} \tilde z_j)} = 0 \, , \quad 
\Im{(\sum_{j=1}^{\Delta m/2} \tilde z_j)} = 
\frac{1}{2}  
\Im{\left( \sum_{k=1}^{m^+} z_k - \sum_{l=m^+ + 1}^{m} 
z_l \right) } \, .
\end{equation}
The set of functions $\cons_2$ given by 
equations (\ref{cons-multi-vortex-1}, \ref{cons-multi-vortex-2}) 
forms an $\Delta m/2$-dimensional vector space.

\end{document}